# Performances of the Italian Seismic Network, 1985-2002: the hidden thing

Alessandro Marchetti, Salvatore Barba, Luigi Cucci, and Mario Pirro

Istituto Nazionale di Geofisica e Vulcanologia, Roma, Italy

**Abstract**

Seismic data users and people managing a sesimic network are both interested in the potentiality of the data, with the difference that the former look at stability, the second at improvements. In this work we measure the performances of the Italian Telemetered Seismic Network in 1985-2002 by defining basic significant parameters and studying their evolution during the years. Then, we deal with the geological methods used to characterise or to plan a seismic station deployment in a few cases. Last, we define the gain of the network as the percentage of located earthquakes with respect to the total recorded earthquakes. By analysing the distribution of non-located ("missed") earthquakes, we suggest possible actions to take in order to increase the gain.

Results show that completeness magnitude is 2.4 in the average over the analysed period, and it can be as low as 2.2 when we consider non-located earthquakes as well. Parameters such as the distance between an earthquake and the closest station, and the RMS location decrease with time, reflecting improvements in the location quality. Methods for geologic and seismological characterisation of a possible station site also proved to be effective. Finally, we represent the number of missed earthquakes at each station, showing that nine stations control more that 50% of all missed earthquakes, and suggesting areas in Italy where the network might be easily improved.

# Introduction

Seismic data analysis at the Istituto Nazionale di Geofisica e Vulcanologia (INGV) begins as soon as the data are acquired, going from automatic processing for very rapid earthquake response to analysts review for earthquake catalogues and quality control. Immediate earthquake information for Italy and surrounding areas includes location, magnitude, expected damage, and, for larger events, source description and historical analysis. INGV then reports earthquake information to public and private agencies.

On a daily basis, analysts revise the previously reported events and read new ones, picking phases, locating the events, and reporting magnitude. In addition to local and regional earthquake location, analysts read teleseismic phases. All this work converges into standard location reports commonly called "seismic bulletins".

Over the past 20 years, the INGV has invested in the development of the software needed for automated, real-time earthquake analysis. Specifically, in the very latest years the station density, the instrumentation quality, and the software capabilities strongly increased; this resulted in greater quality, quantity, and immediacy of earthquake locations. Analogue transmission and chemical paper recordings gave up to digital connections and broadband seismometers, improving the accuracy of readings, whereas the greater station density allows recognising earthquake sources with more details.

The long-lasting dataset allows studying changes of seismicity parameters with time, e.g., b-value, moment release rate, and clustering properties, and the amount of new analyses strictly follows the increase of data quality. The knowledge of human induced changes thus assumes a great importance in planning the analyses and evaluating the results.

Parallel to these improvements in software and instrumentation, in the past 10-12 years a great effort has been made to come up to good standards in the process of site selection through detailed geological surveys. This process is very important, because accounting for the geology of a potential site can improve the station efficiency maximizing the return of the above-described technological improvements.

This work analyses the temporal changes of quantities that depend on the ongoing transformations, e.g., minimum hypocentral distance, completeness magnitude, and RMS among the others. In addition, we also illustrate how the criteria of choosing installation sites changed with time, showing a few examples of some of the most interesting geological situations observed in the field since when the standard investigations at the sites are operating. In order to highlight the changes due to the network, like station density and recording quality, we relocate 1985-2002 earthquakes adopting year-2004 criteria (minimum number of phases, weights, etc.) over the whole dataset. Such information gain interest because of the widespread availability of data due to pagers, e-mail, and the www.

## Data and Instrumentation

### The Italian Telemetered Seismic Network

The Italian Telemetered Seismic Network (ITSN) has been continuously operating short-period instrumentation since the year 1980. During the years, there have been two major developments phases. The first occurred during the years 1981-1984: the network moved to digital acquisition (Taccetti & Mele, 1989; Boschi *et al.*, 1990), and the station density strongly increased (De Simoni, 1987). The second has been occurring since the year 2002: transmission changed from analogue to digital (Badiali and Mele, 2000), including satellite, and new kind of instruments have been deployed, mostly broadband. Figure 1 shows the location of seismic stations in 1984 and in 2002, in between the two largest development phases.

### The Dataset

We dealt with all the local and regional data recorded by the ITSN (Barba *et al.*, 1995a; 1995b) from 1 January 1985 to 31 August 2002 (hereinafter 1985-2002). "All data" means that we did not select any subset a priori. More specifically, we used 1) earthquakes that have been located, 2) earthquakes that had no enough data to give a reliable location, and 3) sparse readings where the S-P time and magnitude are reliable. In all, the specified data set consists of 113199 seismic events, of which 45191 have been reliably located, whereas 68008 are either poorly located or sparse phase readings. Among these latter events, the greatest majority (66698) have reliable S-P time reading at the closest station. These data get together into the public seismic bulletins, available through the www (http://www.ingv.it/) and ftp (ftp://ftp.ingv.it/bollet/).

**Completeness magnitude**

Events in the catalogue have magnitudes between M ~ 1.5 and M ~ 6.0. Completeness magnitude is somewhat troublesome to establish because it depends on area, time, and location quality. We used the maximum of the derivative of frequency magnitude distribution to derive the completeness magnitude for the whole dataset. We computed the completeness magnitude for the whole catalogue (1985-2002), and for two sub-catalogues, 1985-1993 and 1997-2002, respectively. We found M 2.4 for located earthquake, whereas poorly- or not-locatable events are complete from magnitude M 2.2. Only for the years 1997-2002, we find the completeness magnitude for located earthquakes to be M 2.3, thus reflecting the improvement in the network geometry in the latter years.

**On the importance of poorly-located events**

The poorly-located and the not-locatable earthquakes ("missed" earthquakes) constitute the majority of events in the dataset. We call here "poorly-located" the local events defined by the seismic analysts as such. On the other side, "not-locatable" are local earthquakes with seven phase readings or less, that have both P- and S-wave readings at the closest station. If the S-wave is not reliable at the closest station then we completely discard the event from the dataset. Some analysis can still successfully make use of such data if the location accuracy concerns less than magnitude completeness. For example, analysing such data might be very useful in assessing the detection capabilities of the network, or in suggesting improvement for the network geometry.

For sake of simplicity, we group together the poorly-located and the not locatable events, referring to them just as the poorly-located events. We assume as event parameters – in place of origin time and hypocentral coordinates – the arrival time of earliest P-wave, the coordinates, and the S-P time difference, all relative to the closest seismic station.

## Looking inside seismic bulletins

Increasing the seismic station number (from 41 in the year 1984 to more than 100 in year 2002) has produced a huge quantity of low-magnitude recordings. Percentage of magnitude M<= 2.5 events increased from 20% in 1982 up to 62% in 2002. This increase suggests that ITSN data quality improved with time, but it hardly quantifies such improvement. Thus, we study here how the possible indicators of data quality, like the location RMS, the minimum and maximum hypocentral distance, and so on, changed with time. These changes were mostly due to the network geometry and to the site characteristics.

### Variation of minimum and maximum distance with time

In order to understand how the location quality improved with time, we first assume that the error on hypocentral coordinates decreases with the distance between the earthquake and the closest station that recorded it. Here, we are neglecting the influence of the velocity model, and we are assuming that the network geometry might be considered approximately constant around the hypocenter. We selected 36942 local events with minimum hypocentral distance less than 50 km occurred from January 1985 to August 2002. Then we considered monthly averages of the minimum distance and the associated standard deviation. We observe (continuous line in Figure 2) that minimum distance decreased 1.4 km from 24.3 km to 22.9 km in the average on the whole Italian territory. Apparently, this decrement might be surprisingly small, but it is not if we consider that the ITSN also developed in areas with previous little instrumental coverage, and that coastal earthquakes contribute to the average, too. If we consider only areas that were already properly monitored, we find that minimum distance decreases much more. For example, 6 km in the Northern Apennines (5774 events, line "a" in Figure 3), 5 km in the Central Apennines (10866 events, line "b"), and even 15 km in the Matese-Irpinia, Southern Apennines (3306 events, line "c"). In this last case, the Sannio-Matese local seismic network (Di Maro & Marchetti, 1992) was incorporated into the ITSN, lowering the station spacing to 15-20 km. Changing the 50-km cut-off to other distances produces similar results. Therefore, we can conclude that the Network was successfully developed in the well-known hazardous areas, whereas a fair expansion was brought forth in less hazardous (or less known) areas.

As a second important issue, we want to verify how the detectability of an earthquake improves with time. There are two ways to define detectability. At short distances, we say than an earthquake is detectable if it has reliable P- and S-wave arrival time readings at the closest station, so detectability improves if the number of low-magnitude earthquakes with both P- and S-readings increases. It actually improves, increasing the percentage of low-magnitude events from 20% to 62% as above described. The second possible definition concerns the maximum distance at which the earthquake gives reliable P-readings. Many factors affect the so defined detectability: the network density (only at short distances), the quality of the geological sites, the efficiency of transmission lines, the instrument sensitivity, and the reliability of the detection algorithm. We selected 40536 events, with maximum distance less than 600 km, occurred from January 1985 to August 2002, and plotted monthly averages of the maximum distance. We can observe (continuous line in Figure 4) that maximum distance slightly decreases until year 1997, and it slightly increases after the same year. This result suggests that the only intervening factor in 1985-2002 has been the network density, while the other characteristics have been nearly unchanged. We notice that after 2002 great changes occurred in all the other areas, as described in the Network section: digital was used instead of analogue transmission, many broadband instruments were deployed, and digitisation from 16 bit moved to 24 bit. In addition, a strongly improved detection algorithm is running since 2002.

**<u>Variation of RMS with time</u>**

The RMS of travel-time residuals is commonly used to assess the reliability of the earthquake location. This practice suffers of well-known problems, because a minimum of RMS does not insure a realistic solution. A low RMS is easily achieved with few phase readings, whereas a good location often is not. Despite of these limitations, RMS variations can indicate changes in the goodness of location, all the other conditions left constant. In this section, we describe how RMS varies with time, with magnitude and how it depends on the number of P- and S-readings actually used in locating the events.

By representing monthly averages of the RMS travel-time residuals, we can see that RMS decreases with time until year 1997 (continuous and dashed lines in Figure 5). This reduction reflects the increase in station density through the years and the consequent improvement in location accuracy both for all-magnitude events (continuous line in Figure 5) and magnitude M>=3.3 events (dashed line). Since 1997, the RMS was approximately constant, confirming the trend in station number. By the end of 2002, the installed stations increased from 90 to 100.

Magnitude also affects RMS, although not directly. We observe that RMS is nearly constant (RMS $\approx 0.6 \div 0.7$) for events of magnitude M>4. For lower magnitudes, the RMS decreases, probably because of the fewer travel-time data involved in the earthquake location. This result confirms that fewer phases are likely to produce a lower RMS with a possibly worse location accuracy.

**<u>Dependence of RMS on the number of phases</u>**

Finally, we verify how the number of P- and S-wave arrival-time data affect the RMS of the solution. Neglecting the events located with very few data, which avoids fictitious features, we can plot RMS averages versus the number of P- and S-wave arrival times. From Figure 6 we deduce that the minimum of RMS is reached for a number of P phase readings comprised between 15 and 40. For more than 40 P phases, the RMS is nearly constant, showing that additional data do not improve the RMS. The same Figure 6 shows that the optimal number of S phases is 4, and that the use f more than 20 S phases does not improve the RMS. We remember that these are "average" results and that might therefore vary for specific areas or at the borders of the network.

## Geological characterisation of some ITSN station sites

### Methodology for station site characterisation

The phases of development of the ITSN that have been observed during the years find their counterparts in the process of selection of the seismic stations. Indeed part of the improvement in the quality of the "seismological datum" has to be related to more rigorous criteria of site selection. Following an early period in which a site was normally selected by a simple test of the signal with a portable recorder and by quick morphological observations, in 1992 the INGV started a systematic geological and geomorphological survey (Cucci and Pirro 1992, 1993) along with a detailed analysis of the seismic noise at each site. This important work allowed to 1) improve significantly the set of data available for each surveyed station of the ITSN; 2) try to link the geology of one site to possible anomalies in the record of the seismic waves (different signal to noise ratios, capability of recording seismic events, static residuals, etc.); 3) identify the best possible geological conditions at sites where particular needs suggested to move the seismometer; 4) provide useful guidelines to be followed in selecting new sites.

In particular, in this article we will focus our attention on the geological characteristics of the sites and refer to Cimini et al. (1994) and Cucci et al. (1994) for the geophysical aspects of the work.

In theory, the best geological characteristics are those that supply the site with the lowest possible seismic noise. Consequently, the most frequent search is for competent rocks in low relief areas, along-depth continuity of outcrops, lack of fracturing, thin soil cover, absence of active faults and/or slope failure processes in the vicinity of the site. Nevertheless, the geological characterisation is but one of the several features to be assessed when selecting a new site, so that often a favourable geological situation can be in conflict with other important logistic needs such as an easy access to the site or the availability of electricity and phone lines. This explains the complexity and variability of the situations that have been observed while surveying the sites of the ITSN.

At present, we collected geological data regarding more than thirty already existing seismic stations plus a number of sites for future stations. Our working methodology mainly follows two steps. In order to achieve a general overview of the study area we usually collect and analyse geological, hydrogeological and geomorphic data at regional scale (order of tens of kilometres from the site). The processing of these data provides 1:100000 scale maps and sections that allow recognising the most important geological structures. The second step of our work is the detailed geological survey of the area surrounding the site. Each outcrop is briefly described in terms of lithological characteristics, attitude, state of fracturing and possible displacements; further observations concerning the drainage pattern, erosion features and karst phenomena complete the survey. GPS measures provide elevation and coordinates of the site. The data collected in the field are often integrated by aerial photo analyses and by the stratigraphies of wells and boreholes. Geological maps and sections are the final product of the field survey; the scale used for the maps (1:5000) allows to put in evidence single ~20 m-wide outcrops over a ~1 km$^2$ large area. The geological section shows a tentative reconstruction of the bedrock up to 200-300 metres below the site.

At present, we collected geological data regarding about forty existing seismic stations plus a number of sites for future stations. In the following, we show maps and sections of four

ITSN sites that are representative of some of the most interesting geological situations observed in the field.

**Field cases**

Figure 7 shows the results of the survey at the FVI-Forni Avoltri station in the Eastern Alps. This is an example of a site that satisfies many of the above-mentioned requirements and that exhibits scarce geological and/or geomorphological problems. The good geological quality of this location is also confirmed by the seismic noise recorded at the site (Cucci et al. 1994).

Figure 8 shows a geological section at the site RFI, on the volcanic edifice of Roccamonfina in central-southern Italy. Although the choice of this area was mainly suggested by the criterion of maintaining the geometry of the network, the field survey allowed us the selection of a site with considerable thickness of competent rocks by the central latite dome.

Moraines are heterogeneous glacial deposits that usually appear chaotic, colluviated and weakly cemented; these poor geotechical characteristics make moraines not very suitable to host seismic stations. At the site ORO (Oropa in the north-western Alps, see Figure 9, station deployed before 1992), <15 ka, 30-40 m-thick Wurm moraines are deeply eroded by the adjacent stream and poorly cemented. In this case, the geological field survey suggested moving the station to a near site with bedrock directly outcropping, with a consequent improvement of the signal/noise ratio.

The map and section in Figure 10 represent the geological setting at the PII-Pisa station in Central Italy. Here, the stratigraphy of a number of boreholes greatly helped to evaluate the thickness of the alluvial deposits in the area and to choose the most favourable site where to deploy the seismometer. The preferred site exhibits a low level of seismic noise, with a peak at 0.7 Hz (Cimini et al. 1994) possibly due to a site resonance induced by the nearby Arno River.

**From past data to future developments**

In the period from 1 January 2000 to 31 December 2002 (hereinafter 2000-2002) have been recorded 24458 events, 35% of which have been reliably located (the "gain"). Despite of nearly doubling the absolute number of reliably located events, the gain decreased from 44% (1985-1987) down to 35% (2000-2002). This change depend on the network only, being all the parameters used in the location procedure the same. In this section, we investigate the reasons of such earthquake losses, suggesting actions to improve network gains. Such improvements do not seem out of reach – where not yet undertaken, as nine (≈10%) stations record more than 50% of not-localised earthquakes (see Table 1).

Among the 15437 poorly-located events and sparse readings, we select the 15437 with a reliable S-P reading at the closest station. The distribution of S-P delay times is positively skewed, and the central 99% range is 0.73 s-12.34 s. The median of such distribution is 3.91 s, with an estimated standard error of 2.43 s. For each station, we computed the median of the S-P delays relative to non-located earthquakes recorded at the station; then we converted such median to a distance by multiplying the delay by 6.4 km/s. We represent for each station a circle whose radius is the median distance from the station (Figure 11) and whose colour represents the number of non-located events recorded at the station. We can easily identify the areas that might need actions to improve the network gain (although in some cases the action has already taken at the time of writing). From Figure 11, we select the following cases (South to North, counter clockwise):

1) In Northwestern Sicily, the USI station (red colour in Figure 11) recorded more than 1000 events as the closest station (Table 1). Such large number derives from a sequence occurred 40 km offshore NW Sicily in the second half of 2002. Improvements in this case could have come only from offshore seismometers.

2) In, Eastern (Belice Valley) and Central Sicily, despite of the good station coverage, a significant number of earthquakes were missed around the CLTB (cyan) and GIB (green) stations. The low number of S-readings at CLTB might be a starting point for investigating ways to improve station quality in this area. In Southern Calabria, SOI (cyan) seems to indicate a similar problem.

3) In Southeastern Sicily, the existence of only three stations apparently makes several earthquakes to get lost. Therefore, an additional station close to PZI and MEU (cyan) might be enough to add details to the local seismicity.

4) In Southern Apennines, we find cyan and green stations along the Tyrrhenian coast (MRLC, SGO, MGR, SLCN, and CSSN), indicating low efficiency, and several sparse blue stations along the Ionian coast, indicating the presence of local seismicity hard to detail. The former problem might be due to high attenuation in the Tyrrhenian side, whereas the latter to the scarce network coverage. Recently (2003), the station density increased in the area, possibly overcoming the latter problem.

5) In the Central Apennines, stations CPI2 (yellow), SDI, RNI2, RVI2 (green), and PTQR (cyan) indicate low network efficiency in a region with high station density. More to the North, the high number of non-located earthquakes at MNS (yellow), SNTG, ASS, and NRCA (red) indicate low station coverage. In this area, the station density recently increased.

6) In Northern Apennines, the great number of missed earthquakes at CSNT (cyan), CRE, SFI, and PGD (red) indicate a poor network density/geometry, especially at the West of the chain. Increasing the density in that location might solve, at the same time, the geometry problems evidenced by stations BDI, GSCL, ZCCA (cyan).

7) In the Alps, especially around ORO station (blue), there is a certain amount of diffuse seismicity that might easily detailed by increasing the station density. ORO shows how a good quality station allows recordings with S-P reliable readings at significant distances (median distance ~65 km).

**Conclusions**

We analysed performances of Italian seismic network in 1985-2002, i.e., we defined basic significant parameters whose evolution in time might indicate quality of the earthquake locations. Then we verified the gain of the network, defined as the percentage of located earthquakes with respect to the recorded earthquakes, and suggested possible actions to take in order to increase the gain.

Completeness magnitude can be as low as 2.2 when we consider non-located earthquakes, whereas the completeness for located earthquakes is 2.4 in the average, over the analysed period. When the location reliability is not the prime concern, the non-located earthquakes can be used in specific studies, for example to analyse dependence of the b-value on time.

The distance between an earthquake and the closest station decreases ~6 km (~20%) or even ~15 km in specific areas. Such variation are real, i.e. they are measured over real earthquakes in real operating conditions, and not speculative. Therefore, this numbers do not account for density improvement brought forth in areas after a seismic sequence.

RMS of the location decreases regularly with time, for both lower and higher magnitude events, from 0.5 s to 0.4 s, and from 0.7 s to 0.6 s respectively (Figure 5). This values do not account for improvements in the velocity model obtained during the year, as we made calculations by using only one model. Therefore, the RMS reduction only reflects network changes.

Methods for geologic and seismological characterisation of a possible station site also proved to be effective. We relate the good performances of ORO station to the good site quality achieved after a joint geological and seismological survey.

Finally, we represented the number of missed earthquakes at each station. We showed that nine stations control more that 50% of all missed earthquakes, and suggested areas in Italy where the network might be easily improved. The low gain is apparently due to insufficient coverage in some cases and to station or site quality in others.

**Figure Captions**

Figure 1. The Italian Telemetered Seismic Network at the beginning of year 1985 (black triangles) and at the end of year 2002 (black and white triangles).

Figure 2. Monthly averages (dots) of minimum hypocentral distance and associated standard deviation (bars) from year 1985 to August 2002. Only events with minimum distance less than 50 km contributed to the average. The line represents the linear fit of the minimum distance versus time.

Figure 3. Monthly averages (continuous lines) of minimum hypocentral distance like in Figure 2, but relative to (a) Northern Apennines (blue), (b) Central Apennines (red), and (c) Matese-Irpinia in Southern Apennines (green). The dashed lines represent the linear fit of the minimum distance versus time.

Figure 4. Monthly averages (dots) of maximum hypocentral distance and associated standard deviation (bars) from year 1985 to August 2002. Only events with maximum distance less than 600 km contributed to the average. The line represents the linear fit of the maximum distance versus time.

Figure 5. Monthly averages (dots) of RMS travel time residuals and associated standard deviation (bars). The lines represent the linear fit of the RMS versus time. Open dots and dashed line are relative to Magnitude M>=3.3 events, whereas closed dots and continuous line are relative to all events.

Figure 6. Average of location RMS and its associated standard deviation versus the number of P- (dots) and S-wave (crosses) arrival time data actually used in the location. The "sweet spot" for RMS is with a number of P readings comprised between 15 and 40, and a number of S readings of 4. To avoid overlapping bars, the standard deviation bars versus the number of S readings have been moved to the right side of the figure, whereas at the left side only the average values (crosses) are plotted.

Figure 7. Geological map and section at the site of the station FVI (Forni Avoltri). Contour topography isobaths is also shown.

Figure 8. Geological section at the site of the station RFI (Roccamonfina).

Figure 9. Geological map and section at the site of the station ORO (Oropa).

Figure 10. Geological map and section at the site of the station PII (Pisa).

Figure 11. Average S-P distance and number of non-located earthquakes, with the histogram of the no. of stations in bins of 100 events. The circle radius is proportional to the S-P time (50 km radius for 10 s), whereas the colour (blue, cyan, green, yellow, red) depends on the number of data at each station. Circles with more events lie higher. The histogram adopts the same colour scheme of circles, but the vertical scale is cropped at 20 stations for number of events less than 100.

**Table Captions**

Table 1. Median and estimated standard error of S-P delays measured at the closest station for the 15437 not-located earthquakes occurred in 2000-2002. "Events no." is the number of not-located earthquakes, and "cumulate percentage" is relative to the total events.

# Tables

## Table 1

| Num | Station Code | Lat | Lon | <S-P> (s) | Std err (s) | Events No. | Cumulate percentage |
|---|---|---|---|---|---|---|---|
| 1 | SFI | 43.904 | 11.847 | 5.20 | 0.17 | 1811 | 11.73% |
| 2 | NRCA | 42.833 | 13.113 | 3.38 | 0.25 | 1671 | 22.56% |
| 3 | USI | 38.710 | 13.180 | 7.92 | 0.18 | 1046 | 29.33% |
| 4 | ASS | 43.062 | 12.652 | 3.47 | 0.23 | 824 | 34.67% |
| 5 | CPI2 | 41.585 | 14.318 | 3.15 | 0.41 | 618 | 38.67% |
| 6 | SNTG | 43.254 | 12.940 | 4.58 | 2.33 | 547 | 42.22% |
| 7 | MNS | 42.384 | 12.680 | 3.69 | 0.40 | 436 | 45.04% |
| 8 | RGNG | 41.674 | 15.586 | 4.35 | 0.42 | 433 | 47.85% |
| 9 | ARV | 43.498 | 12.944 | 3.74 | 2.58 | 392 | 50.39% |
| 10 | MRLC | 40.761 | 15.489 | 2.91 | 0.45 | 385 | 52.88% |
| 11 | CRE | 43.619 | 11.950 | 2.93 | 0.47 | 375 | 55.31% |
| 12 | DOI | 44.503 | 7.245 | 4.92 | 2.88 | 347 | 57.56% |
| 13 | SDI | 41.709 | 13.810 | 2.98 | 0.72 | 343 | 59.78% |
| 14 | PTCC | 46.405 | 13.353 | 4.31 | 0.68 | 283 | 61.61% |
| 15 | GSCL | 44.350 | 10.587 | 3.78 | 0.53 | 262 | 63.31% |
| 16 | SOI | 38.073 | 16.054 | 5.00 | 1.27 | 251 | 64.93% |
| 17 | BRMO | 46.476 | 10.373 | 2.88 | 0.73 | 234 | 66.45% |
| 18 | MNO | 37.933 | 14.694 | 5.67 | 3.01 | 198 | 67.73% |
| 19 | RNI2 | 41.702 | 14.152 | 3.36 | 1.53 | 178 | 68.89% |
| 20 | PTQR | 42.021 | 13.400 | 4.50 | 0.36 | 171 | 69.99% |

# Table 1

Figure 1

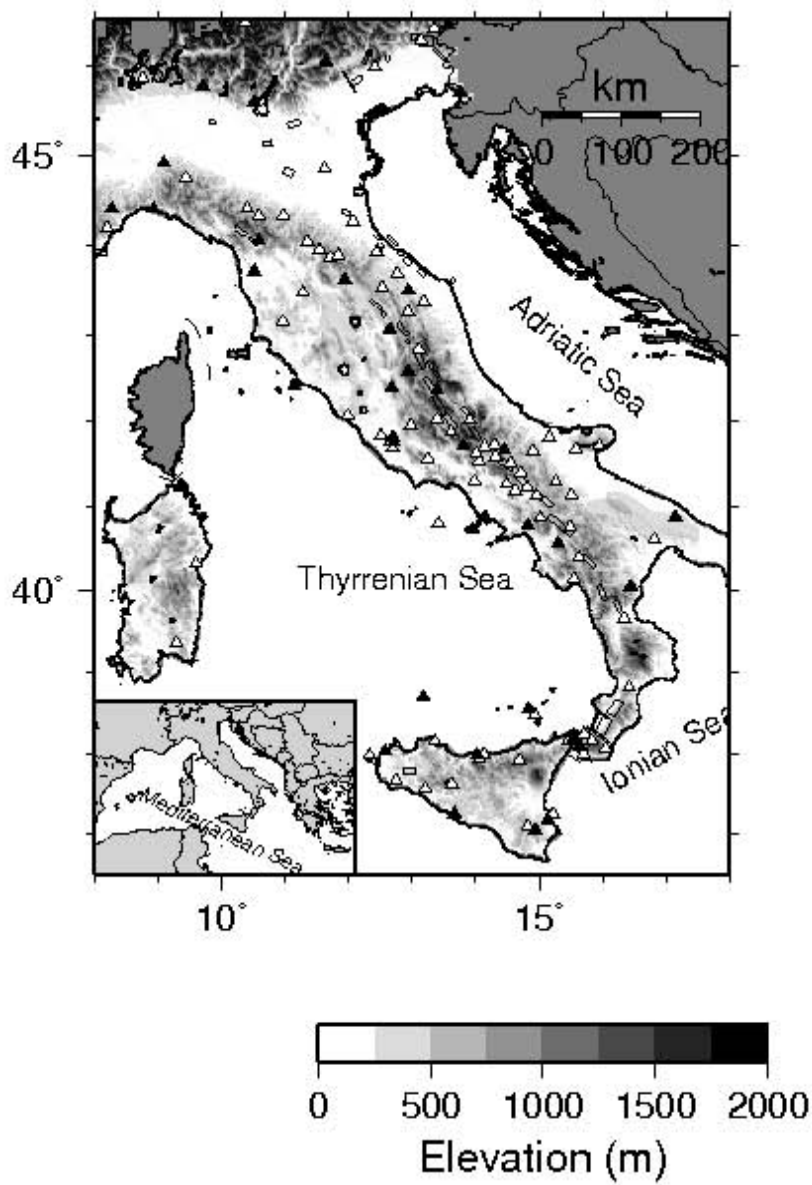

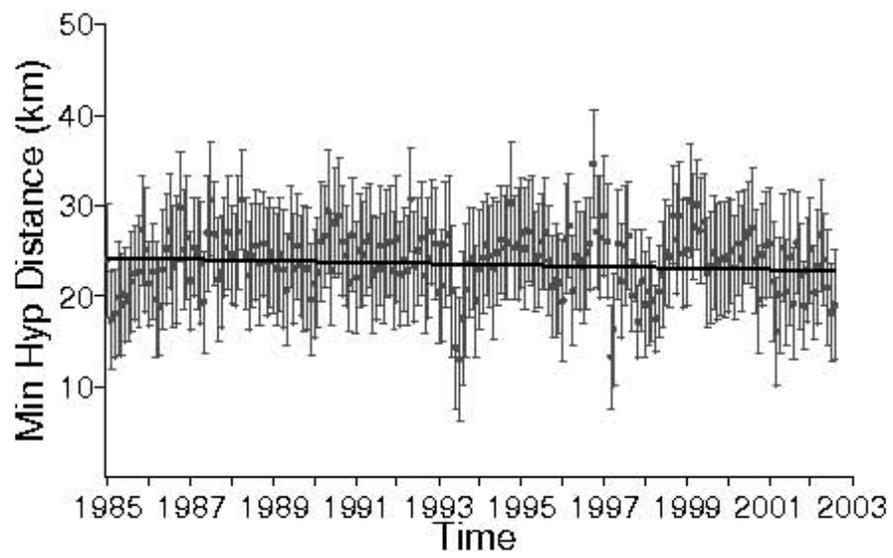

Figure 2

**Figure 3**

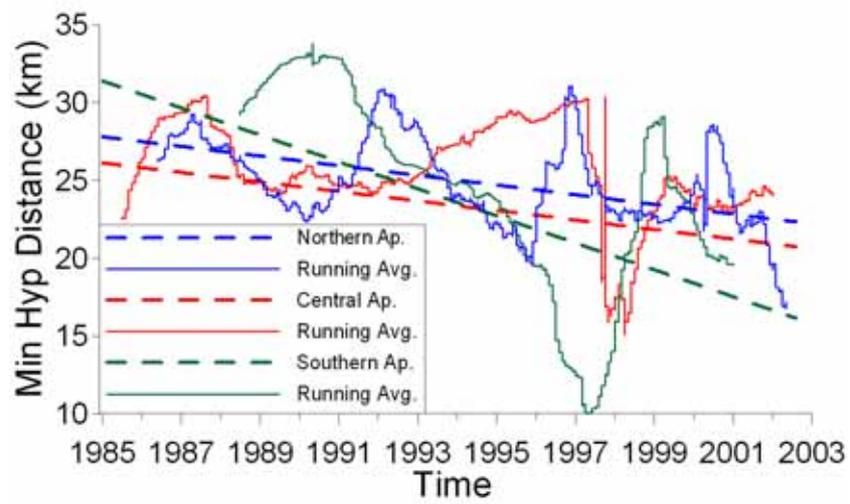

**Fig. 3**

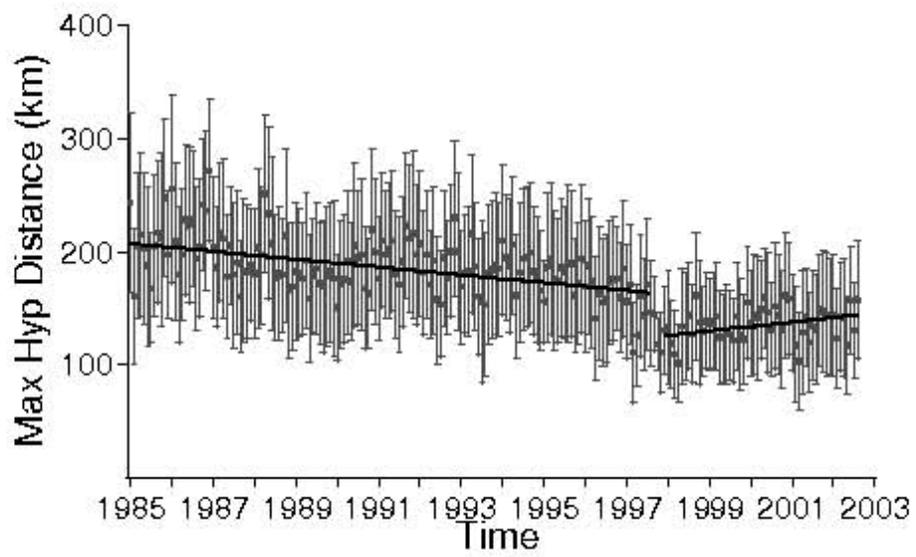

Figure 4

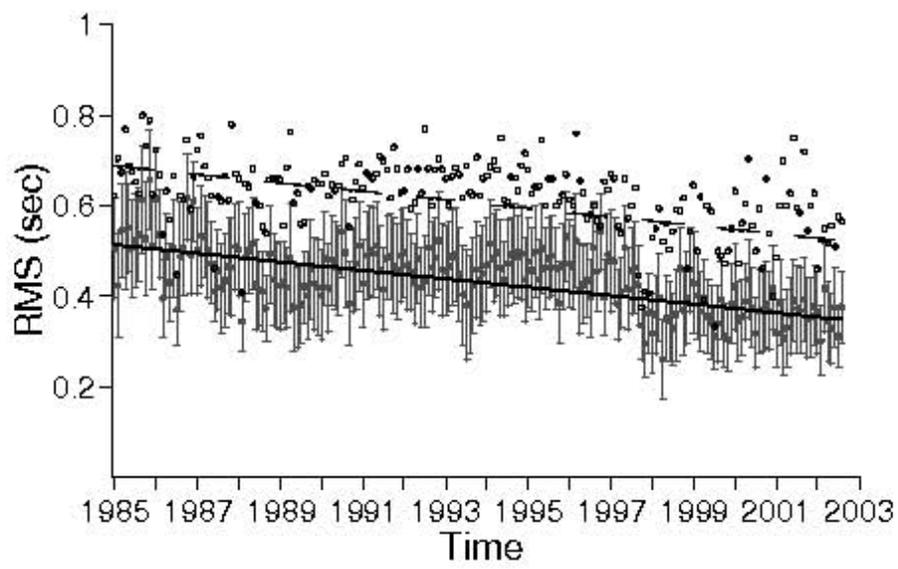

Figure 5

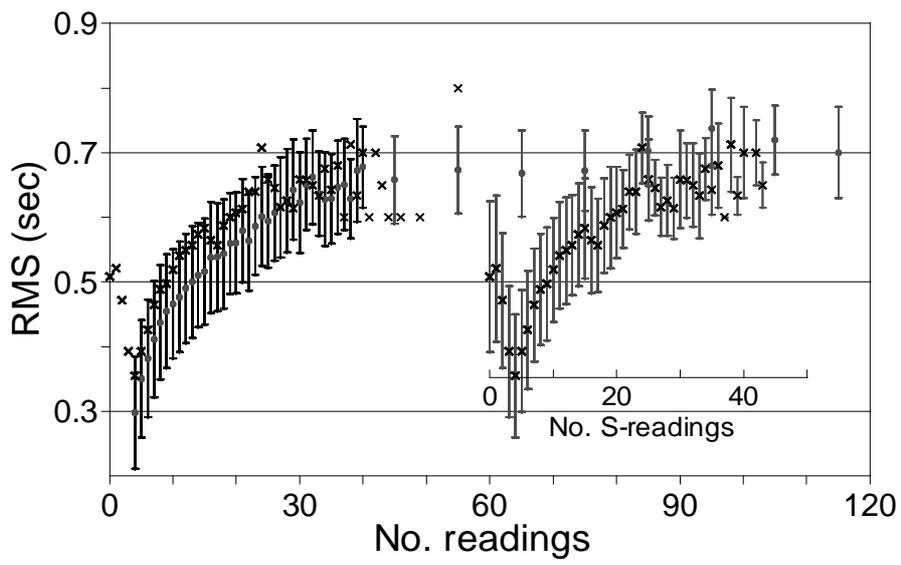

Figure 6

**Figure 7**

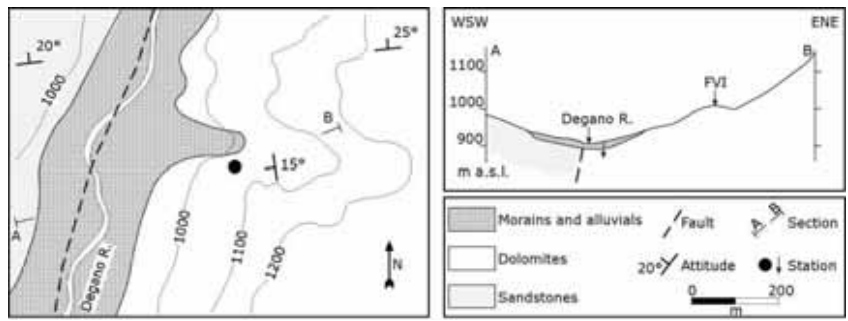

**Fig. 7**

**Figure 8**

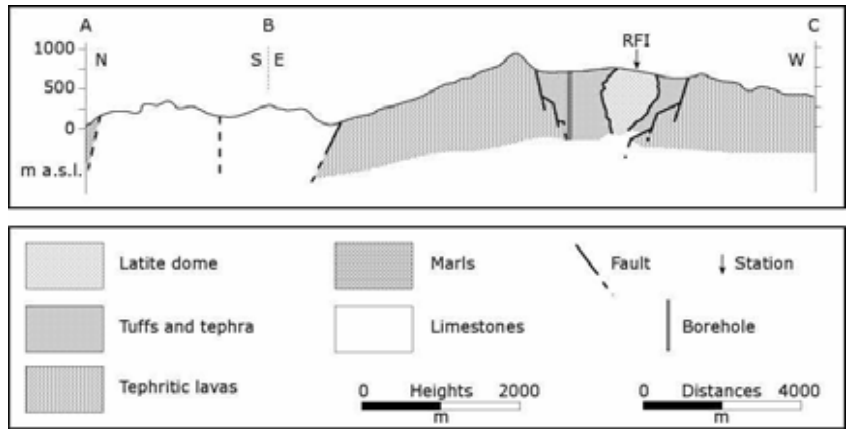

**Figure 9**

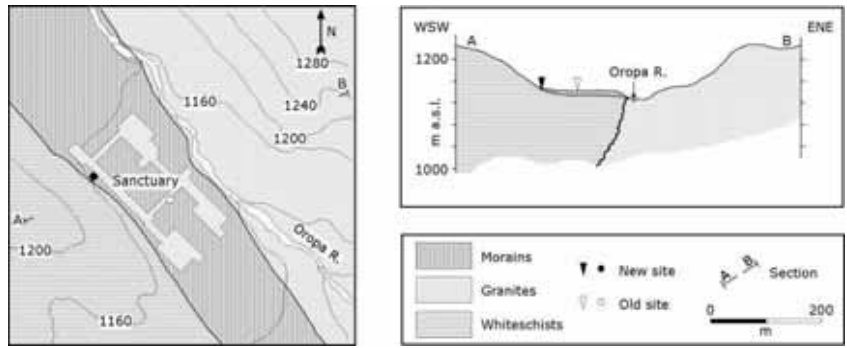

**Figure 10**

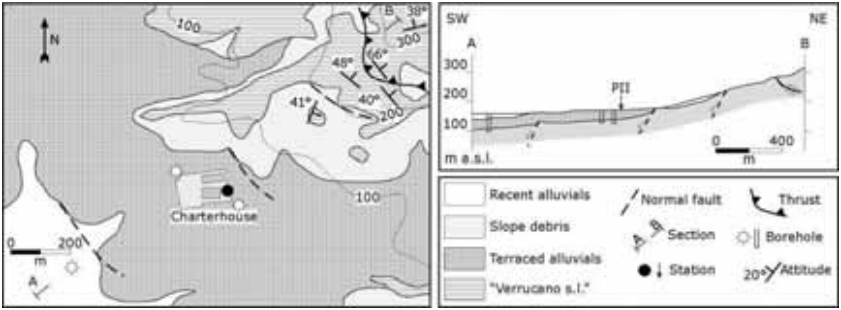

Figure 11

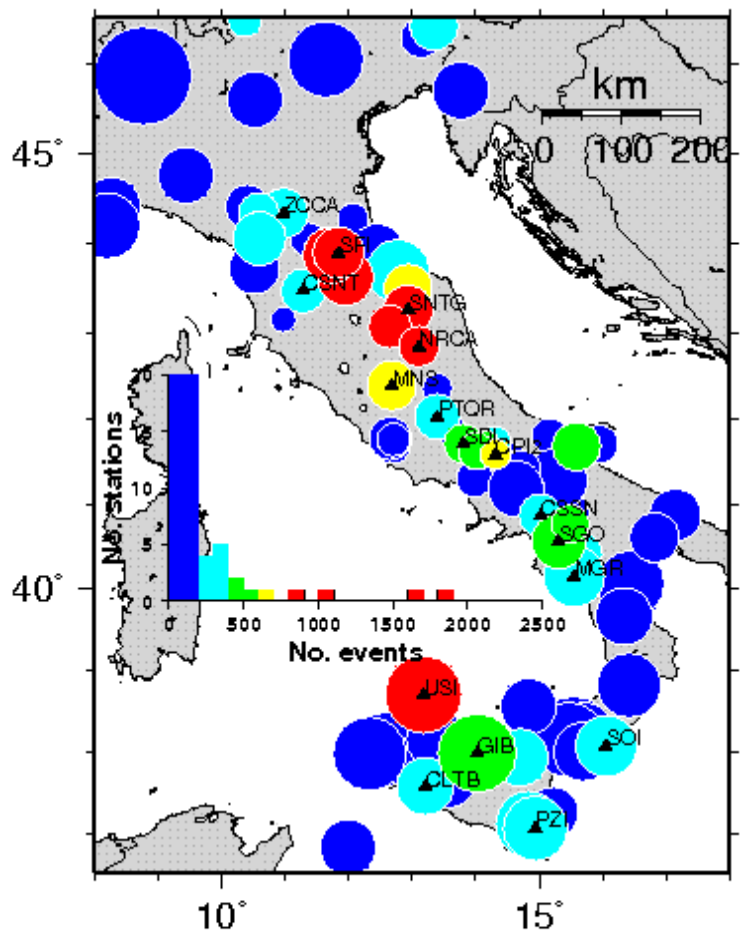